\documentstyle[preprint,tighten,prl,aps,epsf]{revtex}
\begin{document}
 
\title{Phase transition between
           the cholesteric and twist grain boundary C phases}
\author { I. Luk'yanchuk$^{1,2,*}$}
\address{$^1$L.D.Landau Institute for Theoretical Physics, Moscow, Russia.}
\address{ $^2$Departamento de Fisica, Universidade Federal de Minas Gerais, \\
Caixa Postal 702, 30161-970, Belo Horizonte, Minas Gerais, Brazil}

\date{\today }
\maketitle

 \begin{abstract}
\leftskip 54.8pt
\rightskip 54.8pt

The upper critical temperature $T_{c2}$ for the phase transition between the
Cholesteric phase ($N^{*}$) and the Twist Grain Boundary C phase with
the layer inclination tilted to the pitch axis ($TGB_{Ct}$) in thermotropic
liquid crystals is determined by the mean field Chen-Lubensky
approach. We show that the $N^{*}$-$TGB_{Ct}$ phase transition is split in
two with the appearance of either the $TGB_A$ or the $TGB_{2q}$ phase in a narrow
temperature interval below $T_{c2}$. The latter  phase is novel in being
superposed from two degenerate $TGB_{Ct}$ phases with
different (left and right) layers inclinations to the pitch axis.  
\end{abstract}


 %%%%%%%%%%%%%%%%%%%%%%%%%%%%%%%%%%%%%%%%%%%%%%%%%%%%%%%%%%%%%%%%%%%%%%%%%%%%%%%
 
\setlength{\parindent}{5pt} \leftskip -10pt \rightskip 10pt

\section{INTRODUCTION}

A Twist Grain Boundary ($TGB$) state that appears as an intermediate state
at the Cholesteric ($N^{*}$) - Smectic ($Sm$) phase transition in chiral
thermotropic liquid crystals was predicted theoretically by Renn and
Lubensky \cite{First} in 1988 and then, one year later, was independently
observed experimentally \cite{Goodby1,Goodby2}. Since that time a
wealth of properties of this new state were discovered in a number of
experimental \cite{Srajer,Ihn,XrayA,XrayC,Laurence,TGBC,Last,Split,Pres1,Pres2} and theoretical \cite{Second,Renn,MolCryst,Dozov} investigations.

The results of these studies and the results of the present paper are
summarized in the phase diagram of Fig. 1 where the parameters $t,\sigma
_{\bot }$ (whose meanings will be explained in Sect. II) are controlled by the
following experimental conditions: temperature, concentration, pressure etc. The
reason for  such a variety of intermediate phases is that, the direct $N^{*}$%
-$Sm$ transition can not occur in a continuous way since the cholesteric
twist of the director,  ${\bf n}\left( r\right) =(0,\sin \left( k_0x\right)
,\cos \left( k_0x\right) )$, is incompatible with the smectic layered
structure. The last one in chiral liquid crystals is known to be of either $%
SmA$ or  $SmC^{*}$ type: in $SmA$ the director ${\bf n}$ is parallel to the
layers modulation vector ${\bf q}$ whereas in $SmC^{*}$ it is tilted with
respect to ${\bf q}$ by a constant angle $\theta _0$ and forms a conical
precession along the normal to layers. Therefore, the transition occurs
either by the first order untwisting of ${\bf n}\left( r\right) $ or via
formation of intermediate $TGB$ phases. The actual sequence of the
intermediate phases depends on the final $Sm$ state that occurs at low
temperature. There is only one intermediate phase ($TGB_A$) when the
transition goes to $SmA$ \cite{First,Second} and a series of phases when the transition goes to 
$SmC^{*}$ \cite{Renn}.

The general structure of the $TGB$ state is shown in Fig. 2a. The compromise
between the cholesteric twist of ${\bf n}\left( r\right) $ and layered
structure of $SmA,C^{*}$ is achieved by formation of a set of rotated
smectic slabs (blocks), normal to those being follow the pitch ${\bf n}%
\left( r\right) $. The slabs are separated by grain boundaries consisting of
a series of equally spaced screw dislocations that provide the junction of the layers in adjusting slabs. The slab width $l_b$,
dislocation spacing $l_d$, director pitch $P$, and layer spacing $d$ are
related by the following topological constraint \cite{First}:

\begin{equation}
2\pi l_bl_d=Pd  \label{Topol}
\end{equation}
Coupling of the director with the modulation vector ${\bf q}$ results in the
unbending of ${\bf n}\left( r\right) $ close to the block center. When
temperature decreases the director pitch and slab width diverge with further
untwisting of ${\bf n}\left( r\right)$. Finally, they tend to infinity
corresponding to the transition to the $Sm$ state.

The variety of $TGB$ phases in Fig.1 is provided by the different internal
structure of $TGB$ blocks that are shown in Fig.2b-e. Generally, $TGB$
blocks are reminiscent of the final $Sm$ state that occurs at low temperature. 
Therefore, the 
$TGB_A$ block is just the $SmA$ slab, shown in Fig. 2b, confined by grain boundaries, the layers
modulation vector ${\bf q}$ being parallel to the director ${\bf n}$ in the
slab center \cite{First}. Similarly, the $TGB_C$ slab is provided by smectic
layers that are inclined to ${\bf n}$ by the angle $\sim \theta _0$. The
inclination, however, can be done in other ways: when tilted layers are
either parallel to the pith axis $x$ as in Fig. 2b, or tilted to it as in
Fig.2c. We call these phases  $TGB_{Cp}$ and $TGB_{Ct}$. The blocks of the $%
TGB_{Cp}$ and $TGB_{Ct}$ phases actually have the structure of slabs of
differently oriented $SmC$, as distinguished from its chiral analog $SmC^{*}$
by the absence of the director precession. Under certain conditions a
transition $SmC\rightarrow SmC^{*}$ occurs inside the $TGB_C$ slabs \cite
{Renn}. The corresponding $TGB_{C^{*}}$ phase that appears close to the bulk 
$SmC^{*}$ phase (see Fig. 1) will be not considered in this paper.
 
Originally the $TGB_{Cp}$ phase was assumed to be an intermediate $TGB$
state at the $N^{*}$-$SmC^{*}$ transition \cite{Renn} (there it was called $%
TGB_C$). However,  x-ray experiments \cite{XrayC} and  theoretical
estimations \cite{Dozov} demonstrate that the $TGB_{Ct}$ phase is indeed
more stable. In \cite{Dozov} this phase was called as the Melted Grain Boundary
($MGB$) phase to stress that the smectic order parameter vanishes at  grain
boundaries because of the small distance between screw dislocations. We
prefer, however, to use the $TGB_{Ct}$ notation to emphasize the geometrical
structure of this phase.

In this paper we revise the calculation of \cite{Renn} for the {\it upper
critical temperature} $T_{c2}$ ($M_A$-$M_0$-$M_1$-$M_C$ line in Fig. 1 for
the $N^{*}$-$TGB_C$ transition, taking into account the recent proof of the
stability of the $TGB_{Ct}$ phase \cite{XrayC,Dozov} that was not
considered in \cite{Renn}. We confirm and expand the estimation of \cite
{Renn} to the whole region of parameters. In addition we calculate the
principal parameters of the $TGB_{Ct}$ phase: upper critical temperature,
slab width $l_b$ and the  X-ray diffraction pattern that can be measured
experimentally.

Several features that modify the phase diagram calculated in \cite
{MolCryst,Renn} follow from our analysis. The $TGB_A$ phase that was
shown in \cite{First,Second} to be stable when the  transition is to
the $SmA$ phase, penetrates also into the $SmC^{*}$ region as a narrow
stripe in between the $N^{*}$ and $TGB_{Ct}$ phases, and finishes in the
tetracritical point $M_0$ far inside this region. The $N^{*}$-$TGB_{Ct}$
transition is split either by this $TGB_A$ stripe or by the narrow region of
the new $TGB_{2q}$ phase. The $TGB_{2q}$ slab is superposed from two
equivalent $SmC$ populations with left and right layers inclined to the
pitch axis as shown in Fig. 2e.

This new phase can be viewed as a kind of standing density wave quantized by
the grain boundaries. In reality it can be observed in between the 
$N^{*}$-$TGB_A$-$TGB_{Ct}$-$TGB_{2q}$ tetracritical point $M_0$ and the point $M_1$. Location of these points is calculated in this paper. The upper critical
temperature $T_{c2}$ should have a kink at $M_0$. Under certain conditions
the enhancement (oscillation) of $T_{c2}$  between $M_0$ and $M_1$ can be
observed.

A remarkable feature of the $TGB_A$ phase was noted in \cite{First} to be an
analogy with the Abrikosov vortex state in superconductors in a magnetic
field. This completed an analogy between the superconducting transition in
metals and the phase transition between Nematic ($N)$ and  $SmA$ phases in
liquid crystals, first pointed by de Gennes \cite{dG,GP}. In the present
article we show that the $TGB_{Ct}$ phase is the analog of the mixed state in
superconductors with a space-modulated order parameter (like
Larkin-Ovchinnikov-Fulde-Ferrel phases \cite{LO,FF}), providing that
modulation is perpendicular to the magnetic field. Another interesting
analogy we discuss is  the similarity between the $TGB_A$-$TGB_{Ct}$
 transition and the transition between the symmetry differing phases in an
''unconventional'' superconductor $UPt_3$ in the magnetic field \cite{Luk}.

\section{Basic equations}

\subsection{The chiral Chen-Lubensky Model\negthinspace}

On a quantitative level, the appearance of $TGB$ state is described by the
Chen-Lubensky (CL) model \cite{CL}, which is known to be a quite general approach in
explaining  various phase transitions between cholesteric (nematic) phases
and modulated smectic phases. In this model the cholesteric and smectic
phases are described by two coupled order parameters: by the twisted
director $n(r)$ and  the space modulated complex function $\psi (r)$,
where the modulation of the smectic density is given by the real part of $%
\psi (r)$. The resulting energy consists of two parts:

\begin{equation}
{\cal F}_{CL}={\cal F}_\psi +{\cal F}_F  \label{FCL}
\end{equation}
where the elastic Frank energy:

\begin{eqnarray}
{\cal F}_F &=&\frac 12K_1\left( div\,{\bf n}\right) ^2+\frac 12K_2\left( 
{\bf n}\cdot curl\,{\bf n}-k_0\right) ^2  \label{FF} \\
&&\ \ \ +\frac 12K_3\left( {\bf n}\times curl\,{\bf n}\right) ^2  \nonumber
\end{eqnarray}
provides the twisted texture of the director. In the cholesteric phase ${\bf %
n}\left( r\right) =(0,\sin \left( k_0x\right) ,\cos \left( k_0x\right) )$.
The chirality is provided by the parameter $k_0$. When $k_0=0$, expression (%
\ref{FF}) reduces to the elastic energy of the nematic phase with ${\bf n}%
=const$.

The smectic state is described by the Ginzburg-Landau (GL) functional 
${\cal F}_\psi $, which has a finite-$q$ instability for the order parameter 
$\psi (r)$ provided by a gauge derivative ${\bf D}={\bf \nabla }-iq_0{\bf n}$:

\begin{eqnarray}
{\cal F}_\psi &=&a\left( T-T_{NA}\right) \mid \!\psi \mid ^2+\frac 12g\mid
\!\psi \mid ^4+(C_{\Vert }\;n_in_j  \label{FPsi} \\
&&+C_{\perp }(\delta _{ij}-n_in_j))\left( {\bf D}_i\!\psi \right) \left( 
{\bf D}_j\!\psi \right) ^{*}+D\,\left( {\bf D}^2\!\psi \right) \left( {\bf D}%
^2\!\psi \right) ^{*}  \nonumber
\end{eqnarray}
We take the quartic in the gradient term in the isotropic form $D\left( {\bf %
D}^2\!\psi \right) \left( {\bf D}^2\!\psi \right) ^{*}$ which is slightly
different from the original CL model where this term was written as: $%
D_{\perp \,}\left( \delta _{ij}-n_in_j\right) \left( \delta
_{kl}-n_kn_l\right) \left( {\bf D}_i{\bf D}_j\psi \right) \left( {\bf D}_k%
{\bf D}\!_l\psi \right) ^{*}$. This does not change the final results but
simplifies the calculations.

Now we give a brief review of the properties of the CL model. In the
nonchiral case (when $k_0=0$), a second order phase transition from nematic
to smectic phases takes place \cite{CL}. The type of the smectic phase depends on the
sign of $C_{\perp }$ (terms with $C_{\Vert }$ and $D$ are assumed to be
positive). When $C_{\perp }>0$, a transition to the $SmA$ phase with the
order parameter $\psi (r)\sim \psi _0e^{iq_0{\bf nr}}$ occurs at $T=T_{NA}$.
When $C_{\perp }<0$ an additional transversal to ${\bf n}$ modulation
occurs, resulting in a $SmC$ order parameter $\psi (r)\sim \psi _0e^{i{\bf qr%
}}$ with ${\bf q}=(q_0{\bf n,q}_C)$. This modulation is stabilized by the
quartic  gradient term at wave vector $q_C=$ $(-C_{\perp }/2D)^{1/2}$ which
gives the layers inclination angle as $\theta _0=\arctan (q_C/q_0)=\arctan
(-C_{\perp }/2Dq_0^2)^{1/2}$. The $N$-$SmC$ transition occurs at $%
T_{NC}=T_{NA}+C_{\perp }^2/4D$.

In a chiral case \cite {Second} (when $k_0\neq 0$), the situation is more complicated since
the gauge derivatives and anisotropic coefficients in the gradient terms in 
Eq.(\ref{FPsi}) depend on the direction of ${\bf n}$ which is not uniform in a
space. This provides a coupling of the order parameters ${\bf n}(r)\,$ and $%
\psi (r)$ leading to an untwisting of ${\bf n}(r)$ when the smectic layers
are formed. Such a process occurs either directly, by the first order
transition, or via the intermediate $TGB$ state. In both  cases the final
smectic phases are again either $SmA$ or $SmC$.

The scenario of the phase transition depends on the parameters of the CL
model. We formulate those conditions below. When the direct $N^{*}$-$%
SmA,C^{*}$ transition takes place, the critical temperatures are calculated
by comparison of the energies of the $N^{*}$ and $SmA,C^{*}$ phases  \cite{Second}. The $%
N^{*}$-$SmA$-$SmC^{*}$ tricritical point in the 
$(T,C_{\perp })$ plane was found to be: 
\begin{equation}
(T^{*},C_{\perp
}^{*})=(T_{NA}-(gK_2)^{1/2}k_0/a,\;(gK_2^3)^{1/2}k_0/2aK_3q_0^2)
\label{Tri*}
\end{equation}
The transition lines are given by:

\begin{eqnarray}
\;T_{N^{*}A} &=&T^{*},\;C_{\perp }>{C}_{\perp }^{*}
\label{Phasdi*} \\
\;T_{AC^{*}} &=&T_{NA}+(T_{N^{*}A}-T_{NA})\cdot C_{\perp }^{*}/C_{\perp
},\;C_{\perp }<C_{\perp }^{*}  \nonumber \\
\;T_{N^{*}C^{*}} &=&T_{N^{*}A}+(\ C_{\perp }-C_{\perp
}^{*})^2/4aD,\;C_{\perp }<C_{\perp }^{*}  \nonumber
\end{eqnarray}
When the transition occurs via an intermediate $TGB$ state, the critical
temperatures and the detailed structure of the $TGB$ state are provided by a
nonuniform solution of the GL equation obtained from (\ref{FPsi}):

\begin{eqnarray}
&&\ a\left( T-T_{NA}-\frac{C_{\perp }^2}{4D}\right) \psi +g\left| \psi
\right| ^2\psi  \label{GL} \\
\ &=&(C_{\Vert }-C_{\perp })\left( {\bf nD}\right) ^2\psi -D\left( -{\bf D}%
^2+\frac{C_{\perp }}{2D}\right) ^2\psi  \nonumber
\end{eqnarray}
where, $\widehat{A}^2\psi =\widehat{A}\left( \widehat{A}\psi \right) $.

In this paper we are interested in the structure of the $TGB$ state just
below the $N^{*}$-$TGB$ phase transition that takes place at {\it the upper
critical temperature} $T_{c2}$. We consider the case when $C_{\perp }<0$, 
that is when the system has a tendency to form the $SmC^{*}$ phase at low
temperature. In the following subsection we derive the basic equations that
describe this transition.

\subsection{Cholesteric-TGB transition}

The order parameter in the $TGB$ state below $T_{c2}$ is given by a
periodic superposition of rotating blocks equally shifted by a distance $l_b$

\begin{equation}
\psi \left( r\right) ={\sum_m}e^{i\gamma _m}f(x+ml_b)e^{i{\bf q}_{\perp m}%
{\bf r}}  \label{Struct}
\end{equation}
The component of the modulation vector transverse to the pitch  ${\bf q}%
_{\perp m}=q_0\left( 0,\sin \left( k_0ml_b+\varphi \right) ,\cos \left(
k_0ml_b+\varphi \right) \right) $, follows the director twist: ${\bf n}%
(r)=\left( 0,\sin \left( k_0x\right) ,\cos \left( k_0x\right) \right) $ with
phase advance (retardation) $\varphi .$ Factors $e^{i\gamma _m}$ are the
degenerate phason degrees of freedom \cite{First}.

The {\it block profile function }$f(x)$ is localized within the block width $%
l_b$ and, together with the phase $\varphi $, provides the structure of the $%
TGB$ slab. In the $TGB_A$ phase $\varphi =0$ and $f(x)$ is a centered
bell-shape function (which has a Gaussian profile $\exp (-k_0x^2/2q_0)$ at $%
C_{\perp }\gg 0\cite{First}$). The structure of the $TGB_{Cp}$ slab (Fig.
2c) is given by a centered bell-shape function $f(\overline{x})$ and nonzero
advancing (retarding) phase angle $\varphi \simeq \theta _0$ \cite{Renn}
that results in the tilting of the layers parallel to the pitch axis . In
the $TGB_{Ct}$ phase the angle $\varphi $ is equal to zero and $f(x)$ is
modulated in the $x\,$ direction as $\exp (\pm i\theta x)$. This corresponds
to the right (left) transverse inclination of the layers as shown in Fig. 2d.

The profile function $f(x)$ of the slab, located at the origin, and phase $%
\varphi $ are found by solving of Eq.(\ref{GL}) with the substitution $\psi
\left( r\right) =f(x)e^{i{\bf q}_{\perp 0}{\bf r}}\simeq
f(x)e^{iq_0(z+\varphi y{\bf )}}$.

Several simplifications are used in Renn-Lubensky theory. Usually the block
width $l_b$ is much smaller than the cholesteric pitch $P=2\pi /k_0$.
Therefore, on the scale of $l_b$ the twist of ${\bf n}\left( r\right) $ is
minimal and approximately is written as ${\bf n}\left( r\right) \approx
\left( 0,\;k_0x,\;1-\left( k_0x\right) ^2/2\right) $. Next, just below $%
T_{c2}$ the amplitude of $f(x)$ is small and only the terms linear in $f(x)$
are relevant in Eq.(\ref{GL}).

The corresponding linearized equation for $f\left( x\right) $ and $\varphi $
is:

\begin{equation}
a\left( T-T_{NA}-C_{\perp }^2/4D\right) f=-{\cal H}f\text{ \quad where}
\label{Lin}
\end{equation}

\begin{eqnarray*}
{\cal H} &=&D\left( -\partial _x^2+(q_0k_0)^2(x-\varphi /k_0)^2+\frac{%
C_{\perp }}{2D}\right) ^2 \\
&&+(C_{\Vert }-C_{\perp })\frac{q_0^2k_0^4}4(x^2-2x\varphi /k_0)^2
\end{eqnarray*}

This equation has a set of localized eigenstates $f_n(x)$ with a discrete
spectrum of {\it eigentemperatures} $T_n$. The upper critical temperature $T_{c2}$ of the $N^{*}$-$TGB$ transition is provided by the maximal value of $T_n$. The block profile function is given by the corresponding eigenfunction $%
f_{nc2}(x)$.

We assume further that $\varphi =0$, that is, the parallel layers'
inclination according to \cite{Dozov} does not occur. The justification of
this approximation will be presented in Sect. IIID.

It is convenient to use the dimensionless units:

\begin{eqnarray}
\overline{x} &=&q_0x,\;\overline{\partial }_x=q_0^{-1}\partial _x,\;\sigma
_{\Vert }=\frac{C_{\Vert }}{4Dq_0^2},\;\sigma _{\perp }=-\frac{C_{\perp }}{%
2Dq_0^2},  \label{Dimless} \\
t &=&a\left( T-T_{NA}\right) /Dq_0^4\text{,}\;\;b=k_0/q_0,\;\;\overline{%
{\cal H}}={\cal H}/Dq_0^4\qquad  \nonumber
\end{eqnarray}

We discuss first the values of parameters $b$,$\sigma _{\perp }$,$\sigma
_{\Vert }$. Parameter $b$ has a sense of the power of the cholesteric twist.
Usually in chiral liquid crystals the interlayer smectic spacing is much
smaller than cholesteric pitch having $b\ll 1$. The small tilting angle $%
\theta _0$ results in $\sigma _{\perp }=\tan ^2\theta _0\ll 1$. In contrast,
parameter $\sigma _{\Vert }$, which is related to the ratio of the layer
compression elastic constant $C_{\Vert }q_0^2$ to the layer curvature energy 
$Dq_0^4$, is larger than one \cite{Dozov,Tri} and therefore is much
large than $\sigma _{\perp }$.

Neglecting $C_{\perp }$ in comparison with $C_{\Vert }$ in the last term of Eq.(\ref{Lin}), we rewrite it in dimensionless units as:

\begin{equation}
(t-\sigma _{\perp }^2)f=-\overline{{\cal H}}f\qquad \text{where}\qquad
\label{Lindim}
\end{equation}

\[
\overline{{\cal H}}=\left( -\overline{\partial }_x^2+b^2\overline{x}%
^2-\sigma _{\perp }\right) ^2+\sigma _{\Vert }b^4\overline{x}^4 
\]

The order parameter $\psi \left( r\right) $ of the $TGB$ state is
reproduced by substitution of corresponding eigenfunction $f(x)$ into (%
\ref{Struct}). Then, one can calculate the free energy ${\cal F}_{CL}$ (%
\ref{Struct}) of the $TGB$ state as a function of the slab width $l_b$ or,
more conveniently, as a function of the geometrical factor $l_b/l_d$. The
actual value of this ratio is found by minimization of ${\cal F}_{CL}$ with
respect to $l_b/l_d$. It was found \cite{First} that $l_b/l_d$ $\simeq
0.9\varepsilon $ when $C_{\perp }\gg 0$ and the $TGB_A$ phase has a Gaussian
profile in the slab. The factor $\varepsilon $ depends the relative
strengths of cholesteric splay (bend) and the twist elastic energies $%
K_{1,3}/K_2$. It varies from $\varepsilon =0$ when $K_{1,3}/K_2=0$ (the
calculation in this case is reduced to minimizating the Abrikosov factor $%
\beta =\left\langle \psi ^4\right\rangle /\left\langle \psi ^2\right\rangle
^2$), to $\varepsilon \sim 1.5$ when $K_{1,3}/K_2$ is large. We will use
this result to evaluate $l_b/l_d$ in the $TGB_{A,Ct}$ phases when $C_{\perp
}<0$.

\section{Results}

\subsection{General}

In this section we calculate the eigenstates of Eq.($\ref{Lindim}$) that
correspond to the upper critical temperature $t_{c2}$, and discuss the block
structure just below $t_{c2}$ for the different types of $TGB$ state at $%
\sigma _{\perp }>0$.

Note that eigentemperatures $t_n$ (including $t_{c2}$) and eigenfunctions $%
f_n(x)$ are generally  functions of $b$,$\sigma _{\perp }$,$\sigma _{\Vert }$%
. Because of the  scaling properties resulting from Eq.($\ref{Lindim}$):

\begin{eqnarray}
t_n(b,\sigma _{\Vert },\sigma _{\perp }) &=&\sigma _{\perp }^2\cdot
t_n(b/\sigma _{\perp },\sigma _{\Vert },1),  \label{Scal} \\
f_n(\overline{x},b,\sigma _{\Vert },\sigma _{\perp }) &=&f_n(\sigma _{\perp
}^{1/2}\overline{x},b/\sigma _{\perp },\sigma _{\Vert },1).  \nonumber
\end{eqnarray}
parameter $\sigma _{\perp }$ can be excluded if one considers the dependence 
$t_n(b)$ on the rescaled coordinates $b/\sigma _{\perp },t/\sigma _{\perp }^2
$.

To proceede with the diagonalization of Eq.($\ref{Lindim}$) 
for $\sigma _{\perp}>0$, recall first the results of \cite{First,Second} for the
opposite case of $\sigma _{\perp }<0$ when a transition occurs to the $TGB_A$
phase. When $\left| \sigma _{\perp }\right| $ is large enough, the operator $%
\overline{{\cal H}}$ was shown \cite{First} to be truncatable to the more
simple form of the harmonic oscillator $-2\sigma _{\perp }\left( -\overline{%
\partial }_x^2+b^2\overline{x}^2\right) +\sigma _{\perp }^2.$ The lowest
eigenstate gives the Gaussian profile $e^{-b\overline{x}^2}$ of the $TGB_A$
block and the upper critical temperature $t_{c2}=2\sigma _{\perp }b$.
This result can be improved if one considers the residual part of $\overline{%
{\cal H}}$ as a perturbation:

\begin{equation}
t_{c2}=2\sigma _{\perp }b-(1+3\sigma _{\Vert }/4)b^2  \label{Tc2A}
\end{equation}
The second term is smaller than the first one if $-\sigma _{\perp
}/b>0.5+0.38\sigma _{\Vert }$, which is  the condition of applicability of
the approximation. The opposite case of $\sigma _{\perp }/b\sim 0$ will be
discussed later.

Figure 3 shows the result of numerical diagonalization of ($\ref{Lindim}$) for
different $\sigma _{\Vert }$ when $\sigma _{\perp }>0$. We use the $t/\sigma
_{\perp }^2$, $b/\sigma _{\perp }$ coordinates to trace the two highest
eigentemperatures (the maximal one corresponds to $t_{c2}$) as function of $b
$. Note first the concurrence of two eigenstates with close
eigentemperatures $t_{+}$ and $t_{-}$ that give the upper critical
temperature $t_{c2}=\max (t_{+},t_{-})$. These levels often cross when $b$
changes, resulting in the oscillating behavior of $t_{c2}(b)$. The
oscillations of $t_{c2}(b)$ are strong at $\sigma _{\Vert }=0$ and more
pronounced still at large values of $\sigma _{\Vert }$.

The appearance of these nearly degenerate eigenstates is related to two
equivalent, right and left inclinations of layers in the $TGB_{Ct}$ slab.
The corresponding block profile functions, $f_R(\overline{x})$ and $f_L(%
\overline{x})$, however, can not be the eigenfunctions of Eq.(\ref{Lindim})
since they do not possess the definite parity with respect to the symmetry
operation $x\rightarrow -x$, which is a property of $\overline{{\cal H}}$.
The proper eigenstates $f_{\pm }$ with eigentemperatures $t_{\pm }$ are
constructed as superpositions of both  populations as $f_{\pm }=f_R\pm $ $f_L
$ that results in the $TGB_{2q}$ phase just below $t_{c2}.$ The function $%
f_{\pm }(x)$ has $\cos $- ($\sin $- ) like oscillations and can be viewed as
a standing wave between the grain boundaries. The effects of commensurability of
this wave with block width $l_b$ stabilize either the even $\cos $-like or
the odd $\sin $-like behavior as the lowest state to have a vanishing order
parameter at the grain boundaries. The block width $l_b$ changes as a
function of $b$ that alternates the order of the $f_{\pm }\ $eigenstates
leading to the oscillations in $t_{c2}(b)$.

The tendency to $SmC$ slab formation at lower temperatures results in the
further transition from the $TGB_{2q}$ to $TGB_{Ct}$ phase determined by the
lowest from $t_{+},t_{-}$ critical temperatures renormalized by the
nonlinear term $g\left| \psi \right| ^2\psi $ in Eq.(\ref{GL}). Below this
second transition, one of the populations starts to be suppressed and a block
profile function is constructed now from  both the $f_{+}$ and $f_{-}$
eigenstates to form the $TGB_{Cp}$ phase with either a $f_L$ or $f_R$
profile of the slab.

Therefore, we conclude that the $N^{*}$-$TGB_{Ct}$ transition is always
split by the intermediate $TGB_{2q}$ phase. At high $b$ the period of
oscillations of $f_{\pm }(x)$ becomes larger than its localization length,
and the lowest eigenstate $f_{+}(x)$ has the single-peak profile of the $%
TGB_A$ phase. The $N^{*}$-$TGB_A$-$TGB_{Ct}$-$TGB_{2q}$ tetracritical point $%
M_0$ corresponds therefore to the highest in $b$ intersection of $t_{\pm }$
eigentemperatures. As  follows from Fig. 3, the appearance of the $TGB_{2q}$
phase becomes practically invisible when $b$ decreases below the second
intersection of $t_{\pm }$ at the point $M_1$. We expect, therefore, that
only the odd $TGB_{2q}$ phase between $M_0$ and $M_1$ can be observed in
reality.

Equation ($\ref{Lindim}$) can be solved analytically in the two limit cases $%
\sigma _{\Vert }=0$ and $\sigma _{\Vert }\gg 1$, when the operator $%
\overline{{\cal H}}$ can be truncated to a more simple form. Although the
case $\sigma _{\Vert }=0$ does not correspond to the real situation of $%
\sigma _{\Vert }>1$ we consider it first since it clarifies the qualitative
structure of the phase diagram that conserves also at large $\sigma _{\Vert }
$.

\subsection{ Case $\sigma _{\Vert }=0$}

When $\sigma _{\Vert }=0$, operator $\overline{{\cal H}}$ is a polynomial 
Schr\"odinger operator for the harmonic oscillator: 
\begin{equation}
\overline{{\cal H}}=\sigma _{\perp }^2-2\sigma _{\perp }\left( -\overline{%
\partial }_x^2+b^2\overline{x}^2\right) +\left( -\overline{\partial }_x^2+b^2%
\overline{x}^2\right) ^2  \label{Polynom}
\end{equation}
Therefore, it has the same set of oscillator eigenfunctions:

\begin{equation}
f_n(\overline{x})=H_n(\sqrt{b}\overline{x})e^{-\overline{x}^2b/2}
\label{Herm}
\end{equation}
where $H_n$ are the Hermitian polynomials. The corresponding
eigentemperatures are given by the equation:

\begin{equation}
t_n=2\sigma _{\perp }\left( 2n+1\right) b-\left( 2n+1\right) ^2b^2
\label{tc2FF}
\end{equation}
The oscillations of $t_{c2}(b)=\max t_n(b)$ (Fig.3a) are related to the
quantum number $n_{c2}$ of the lowest eigenstate that changes with $b$,
unlike what occurs in the harmonic oscillator where the $n=0$ eigenlevel is
always the lowest one.

{\it When} $b>\sigma _{\perp }/2$, the $n=0$ eigenlevel does correspond to
the upper critical temperature $t_{c2}=2\sigma _{\perp }b-b^2$. The block
profile function has a Gaussian shape $e^{-\overline{x}^2b/2}$ in the $TGB_A$
phase. The slab width is calculated as in \cite{First}:

\begin{equation}
\overline{l}_b=2.2\varepsilon ^{1/2}/b^{1/2}  \label{Lba}
\end{equation}

{\it When }$b<\sigma _{\perp }/2$, $n_{c2}$ is given by $\sigma _{\perp
}/2b-1/2$ rounded to the nearest integer. The oscillating upper critical
temperature $t_{c2}(b)$ tends to $\sigma _{\perp }^2$ when $b$ vanishes as
shown in Fig. 3a. The slab width is given by the width of the polynomial $%
H_{nc2}(\sqrt{b}\overline{x})$ as:

\begin{equation}
\overline{l}_b=\sigma _{\perp }^{1/2}/b  \label{Lbb}
\end{equation}
The block profile function $f_{nc2}(\overline{x})$ oscillates as $\cos (\lambda 
\overline{x}+\pi n_{c2})$ with  period $\lambda =2\pi /\left( 2n_{c2}b\right) ^{1/2}$ $%
\simeq 2\pi /\sigma _{\perp }^{1/2}$. This corresponds to the slab of the 
$TGB_{2q}$ phase which is the superposition of two $SmC$ slabs with layers
inclined at $\theta _0=\pm \sigma _{\perp }^{1/2}$. The parity of $f_n(%
\overline{x})$ alternates with $n$ resulting in the different parity of $%
f_{n_{c2}}$ and $f_{n_{c2}\pm 1}\,$eigenstates. So, the admixture of $%
if_{n_{c2}\pm 1}$ eigenfunctions with the $TGB_{2q}$ profile function $%
f_{n_{c2}}$ that occurs at some critical temperature below $t_{c2}$
corresponds to the $TGB_{2q}$-$TGB_{Ct}$ transition.

The tetracritical point $M_0$, where $N^{*}$-$TGB_A$-$TGB_{Ct}$-$TGB_{Ct}$
phases meet is given by $b_0=\sigma _{\perp }/2$, $t_0=3\sigma _{\perp }^2/4$

\subsection{Case $\sigma _{\Vert }\gg 1$}

In the limit $\sigma _{\Vert }\gg 1$ one can neglect $b^2\overline{x}^2$ in (%
\ref{Lindim}) and rewrite $\overline{{\cal H}}$ as:

\begin{equation}
\overline{{\cal H}}=\left( \overline{\partial }_x^2+\sigma _{\perp }\right)
^2+\sigma _{\Vert }b^4\overline{x}^4  \label{Linform}
\end{equation}
The eigentemperatures and eigenfunctions of Eq.(\ref{Linform}) have in
additional to (\ref{Scal}) the scaling properties:

\begin{eqnarray}
t_n(b,\sigma _{\Vert },\sigma _{\perp }) &=&\sigma _{\perp }^2\cdot
t_{c2}(b\sigma _{\Vert }^{1/4}/\sigma _{\perp },1,1),  \label{Scallim} \\
f_n(\overline{x},b,\sigma _{\Vert },\sigma _{\perp }) &=&f_n(\sigma _{\perp
}^{1/2}\overline{x},b\sigma _{\Vert }^{1/4}/\sigma _{\perp },1,1).  \nonumber
\end{eqnarray}
So, $t_{c2}(b,\sigma _{\Vert },\sigma _{\perp })$ at large $\sigma _{\Vert }$
is defined by the universal function $t_{c2}(b,1,1)$. We  therefore use the
scaled coordinates $b\sigma _{\Vert }^{1/4}/\sigma _{\perp }$, $t/\sigma
_{\perp }^2$  for large $\sigma _{\Vert }$ to trace $t_{c2}(b)$ that is
obtained from numerical diagonalization of (\ref{Linform}) (Fig. 3c).

An important conclusion that followes from Fig. 3c is that the phase diagram
for large $\sigma _{\Vert }$ posseses the same features as for $\sigma
_{\Vert }=0$. It includes the domains of the $TGB_A$ and $TGB_{2q}$ phases
that precede the transition to the $TGB_{Ct}$ phase. The $TGB_A$ phase
exists above the tetracritical point $M_0$ and the $TGB_{2q}\,$ phase is
important in between the points $M_0$ and $M_1$. From  Fig. 3c one locates
the points $M_0$, $M_1$ at $b\sigma _{\Vert }^{1/4}/\sigma _{\perp }\simeq
0.36$ and $b\sigma _{\Vert }^{1/4}/\sigma _{\perp }\simeq 0.19$.

Consider now the transition to the $TGB_C$ phase at low field $b\ll
0.36\sigma _{\perp }/\sigma _{\Vert }^{1/4}$ (i.e., below $M_0$) and to the $%
TGB_A$ phase at high field $b\gg 0.36\sigma _{\perp }/\sigma _{\Vert }^{1/4}$
(i.e., above $M_0$). Both the asymptotics for $t_{c2}(b)$ are shown in Fig.
3c by the dashed lines.

{\it Low fields}${\it :}$ $b\ll 0.36\sigma _{\perp }/\sigma _{\Vert }^{1/4}$

At low field when the block width $\overline{l}_b$ diverges, one can assume
that blocks have the structure of $SmC$ slabs where the inclination of
layers to the pitch axis is equal to its bulk value $\theta _0=\pm \arctan
\sigma _{\perp }^{1/2}$ . Then the block's profile function $f(\overline{x})$
has a high-frequency modulation $e^{\pm i\sigma _{\perp }^{1/2}\overline{x}}$
where $+$ and $-$ signs correspond to right and left tilting respectively.
This factor can be excluded by shift: $\overline{\partial }_x\rightarrow 
\overline{\partial }_x\pm i\sigma _{\perp }^{1/2}$ in Eq.(\ref{Linform}). Next,
we neglect $\overline{\partial }_x^2$ in comparison with $2\sigma _{\perp
}^{1/2}\overline{\partial }_x$ (justification of this approximation will be
given below) and write $\overline{{\cal H}}$ as:

\begin{equation}
\quad \overline{{\cal H}}=-\left( 2\sigma _{\perp }^{1/2}\overline{\partial }%
_x\right) ^2+\sigma _{\Vert }b^4\overline{x}^4  \label{paral}
\end{equation}
Rescaling $\overline{x}$ as: $\zeta =(b^4\sigma _{\Vert }/4\sigma _{\perp
})^{1/6}\overline{x}$ simplifies (\ref{paral}) to an unharmonic oscillator
operator: 
\begin{equation}
\overline{{\cal H}}=\left( 16\sigma _{\perp }^2\sigma _{\Vert }\right)
^{1/3}b^{4/3}\left( -\partial _\zeta ^2+\zeta ^4\right)   \label{sdg}
\end{equation}
Finally, $t_{c2}(b)$ and the corresponding eigenfunction of (\ref{paral})
are given by:

\begin{eqnarray}
t_{c2} &=&\sigma _{\perp }^2-1.06\cdot \left( 16\sigma _{\perp }^2\sigma
_{\Vert }\right) ^{1/3}b^{4/3}\   \label{tc2par} \\
f_{c2}\left( \overline{x}\right)  &=&e^{\pm i\sigma _{\perp }^{1/2}\overline{%
x}}\cdot g_1\left( (b^4\sigma _{\Vert }/4\sigma _{\perp })^{1/6}\overline{x}%
\right)   \nonumber
\end{eqnarray}
where $1.06$ and $g_1\left( \zeta \right) $ are the lowest eigenvalue and
eigenfunction of the unharmonic oscillator equation: $(-\partial _\zeta
^2+\zeta ^4)g_1=1.06g_1$. The block width $\overline{l}_b$ is estimated as
the characteristic width $\sim (4\sigma _{\perp }/b^4\sigma _{\Vert })^{1/6}$
of the function $g_1$. More accurately, we can calculate $\overline{l}_b$ in
the same way as  was done in \cite{First} for the $TGB_A$ phase, using the
suitable Gaussian approximation for $g_1\left( \zeta \right) $ as $%
e^{-0.4\zeta ^2}$. Finally we get:

\begin{equation}
\overline{l}_b=3.2\varepsilon ^{1/2}(\sigma _{\perp }/b^4\sigma _{\Vert
})^{1/6}  \label{Lbc}
\end{equation}

At first glance, the eigenstate (\ref{tc2par}) is doubly degenerate with
respect to a sign change in the modulation phase $e^{\pm i\sigma _{\perp
}^{1/2}\overline{x}}$. However, this is an artifact of the approximation:
the term $\overline{\partial }_x^2$ we are neglecting lifts the degeneracy
and splits the transition onto two. In particular, in between the points $M_0
$ and $M_1$ (Fig3c) $\overline{\partial }_x^2$ favors the odd $TGB_{2q}$
phase given by the linear combination ($e^{i\sigma _{\perp }^{1/2}\overline{x%
}}-e^{-i\sigma _{\perp }^{1/2}\overline{x}})\cdot g_1=2ig_1\sin \sigma
_{\perp }^{1/2}\overline{x}$. The $TGB_{Ct}$ profile appears by means of an
additional transition when the even combination $\sim 2ig_1\cos \sigma
_{\perp }^{1/2}\overline{x}$ admixes.

This effect, however, is as tiny (and negligible below $M_1$) as $%
\left\langle \overline{\partial }_x^2\right\rangle \simeq 1/\overline{l}_b^2$
is smaller than $\left\langle 2\sigma _{\perp }^{1/2}\overline{\partial }%
_x\right\rangle \simeq 2\sigma _{\perp }^{1/2}/\overline{l}_b$, where $%
\overline{l}_b$ is given by Eq.(\ref{Lbc}) and $b\ll 0.36\sigma _{\perp
}/\sigma _{\Vert }^{1/4}$. The above estimation confirms the selfconsistency
of the approximation we made. Condition $b\ll 0.36\sigma _{\perp }/\sigma
_{\Vert }^{1/4}$ is equivalent to the requirement that the wavelength of the
transversal modulation $\lambda \sim \sigma _{\perp }^{-1/2}$ is smaller
than the  block width $\overline{l}_b$ given by (\ref{Lbc}). This clarifies
the physical meaning of the approximation.

{\it High fields:} $b\gg 0.36\sigma _{\perp }/\sigma _{\Vert }^{1/4}$

Neglecting now $\sigma _{\perp }$ in comparison with $\overline{\partial }%
_x^2$ we simplify $\overline{{\cal H}}$ as:

\begin{equation}
\overline{{\cal H}}\simeq \overline{\partial }_x^4+\sigma _{\Vert }b^4%
\overline{x}^4  \label{HighF}
\end{equation}
Then  $t_{c2}$ and the corresponding eigenfunction are given by:

\begin{equation}
t_{c2}\simeq -1.39\cdot \sigma _{\Vert }^{1/2}\,b^2\qquad f_{c2}\left( 
\overline{x}\right) =g_2\left( \sigma _{\Vert }^{1/8}b^{1/2}\overline{x}%
\right)   \label{tc2hf}
\end{equation}
where $1.39$ and $g_2\left( \eta \right) $ are the lowest eigenvalue and
eigenfunction of the operator $\partial _\eta ^4+\eta ^4$. Unlike the low
field case, the eigenstate (\ref{HighF}) is nondegenerate and described by
the even bell-shaped function $g_2$, the lowest odd eigenstate being well
separated from $g_2$ as shown in Fig. 3c.

The eigenfunction $g_2\left( \sigma _{\Vert }^{1/8}b^{1/2}\overline{x}%
\right) $ gives the $TGB_A$ - like profile of the block. The block width $%
\overline{l}_b$ is calculated using the results of \cite{First} by noting
that $g_2\left( \eta \right) \sim e^{-0.55\eta ^2}$:

\begin{equation}
\overline{l}_b=2.1\varepsilon ^{1/2}/\sigma _{\Vert }^{1/8}b^{1/2}
\label{Lbd}
\end{equation}

The approximation we made is self-consistent since $\left\langle \overline{%
\partial }_x^2\right\rangle \simeq 1/\overline{l}_b^2$ is indeed greater
than $\sigma _{\perp }$ when $b\gg 0.36\sigma _{\perp }/\sigma _{\Vert
}^{1/4}$. Considering the truncated term $\widehat{h}=2\sigma _{\perp }%
\overline{\partial }_x^2+\sigma _{\perp }^2$ as a perturbation, one improves
the asymptotic for $t_{c2}(b)$. The correction to $t_{c2}$ is given by $%
\left\langle g_2\left| \widehat{h}\right| g_2\right\rangle \simeq 2\gamma
\sigma _{\Vert }^{1/4}\sigma _{\perp }b+\sigma _{\perp }^2$ where $\gamma
=\left\langle g_2\left| \partial _\eta ^2\right| g_2\right\rangle \simeq 0.5$%
. Finally, we get:

\begin{eqnarray}
t_{c2} &\simeq &\sigma _{\Vert }^{1/4}\,\sigma _{\perp }b-1.39\sigma _{\Vert
}^{1/2}\,b^2\   \label{Ass2} \\
\  &=&-1.39\cdot (\sigma _{\Vert }^{1/4}\,b-0.36\sigma _{\perp
})^2+0.18\sigma _{\perp }^2  \nonumber
\end{eqnarray}
In fact, the asymptotic (\ref{Ass2}) is valid also in the region of  small
negative $\sigma _{\perp }/b$. It makes the analogous expressions in \cite
{Second,Renn} more precise and explicitly determines all the
coefficients.

\subsection{Stability of the $TGB_{Ct}$ state.}

Having calculated the transition temperature and structure of the $TGB_{Ct}$
phase we discuss now its stability, namely, whether this phase has the
highest upper critical temperature $t_{c2}$ (or, alternatively, the highest
upper critical field, $b_{c2}$) among other $TGB$ states and under which
condition a transition from $N^{*}$ to $SmC^{*}$ occurs via the intermediate 
$TGB_{Ct}$ phase rather than via the direct first order transition.

To answer the first question, consider the competing $TGB_{Ct}$ and $TGB_{Cp}
$ phases that can be formed at the $N^{*}$-$SmC^{*}$ transition. Dozov's
estimation of the ratio of their upper critical fields \cite{Dozov}, gives
the factor $2\left( C_{\Vert }/2Dq_0^2\right) ^{1/4}=2(2\sigma _{\Vert
})^{1/4}$ which is larger than one as long as $\sigma _{\Vert }>1$.
Therefore, $b_{c2}(TGB_{Ct})>b_{c2}(TGB_{Cp})$ and the $TGB_{Ct}$ phase is
indeed more preferable.

Our calculations are in agreement with Dozov's estimation. To show this,
compare $t_{c2}(TGB_{Ct})$ given by Eq.(\ref{tc2par}) with $t_{c2}(TGB_{Cp})$
calculated in \cite{Renn}:

\begin{equation}
t_{c2}(TGB_{Cp})\simeq \sigma _{\perp }^2-1.06\cdot \left( 1+\sigma _{\Vert
}\right) ^{1/3}\left( 16\sigma _{\perp }^2\sigma _{\Vert }\right)
^{1/3}b^{4/3}  \label{tpar}
\end{equation}
(We recalculated the numerical factor $1.06$). The ratio of the upper
critical fields for the $TGB_{Ct}$ and $TGB_{Cp}$ phases (at given
temperature) is expressed as $\left( 1+\sigma _{\Vert }\right) ^{1/4}$,
which is larger than one. This is consistent with Dozov's estimation when $%
\sigma _{\Vert }\gg 1$, but gives a more precise value of the numerical
factor. Although the starting GL functional (\ref{FPsi}) is slightly
different from that, used in \cite{Renn} (see our remark after Eq.(\ref{FPsi}))
, this does not change the final result: one can show that in a relevant
limit, $\sigma _{\Vert }\gg \sigma _{\perp }$, expressions for $%
b_{c2}(TGB_{Ct})$ calculated for both the functionals are the same. Note
that standard perturbational analysis proves that the $TGB_{Ct}$ and $TGB_A$
phases are stable with respect to   small parallel tilting of the layers
when angle $\varphi $ in Eq.(\ref{Lin}) becomes nonzero.

To answer the second question, note that transition via an intermediate $TGB$
state is not preempted by the direct $N^{*}$-$Sm$ transition when the upper
critical temperature $t_{c2}$ for the $TGB$ state is larger than the
thermodynamical temperature $t_c$ of the direct transition. Recall first the
calculations for the $TGB_A$ phase when $\sigma _{\perp }<0$ \cite{First}.
Comparison of $t_{c2}$ given by Eq.(\ref{Tc2A}) (without $(1+3\sigma _{\Vert
}/4)b^2$) with the $N^{*}$-$SmA$ transition temperature $T_{N^*A}$ (\ref{Phasdi*}),
which in dimensionless units is written as:

\begin{equation}
t_c=(gK_2)^{1/2}b/Dq_0^3  \label{TcA}
\end{equation}
gives the following criterion for stability of the $TGB_A$ phase:

\begin{equation}
\kappa _2>1/\sqrt{2}\quad \text{where}\quad \kappa _2=(gK_2/2\sigma _{\perp
}^2)^{1/2}/2Dq_0^3  \label{CritA}
\end{equation}

When $\sigma _{\perp }>0$ the $t_{c2}$ transition temperature for $TGB_{Ct}$
is given by Eq.(\ref{tc2par}), and the $N^{*}$-$SmC^{*}$ transition temperature $T_{N^*C^*}$
(\ref{Phasdi*}) in dimensionless units is written as:

\begin{equation}
t_c\simeq \sigma _{\perp }^2-\frac{(gK_2)^{1/2}}{Dq_0^3}(1-\sigma _{\perp
}K_2/2K_3)b  \label{Tc}
\end{equation}

The criterion $t_{c2}>t_c$ is always satisfied when $b$ is small enough,
since at small $b$, $t_c$ is linear in $b$ and $t_{c2}$ is proportional to $%
b^{4/3}$. So, strictly speaking, one can always obtain the $TGB_{Ct}$ phase
by preparation of the nearly racemic binary mixture of the left and right
chiral molecules. The above consideration is valid only if $t_c(b)$ has a
positive slope, that is, when $\sigma _{\perp }<2K_3/K_2$. This condition was
used in \cite{Renn,MolCryst} as a criterion for stability of the $%
TGB_C$ state. It is not very likely, however, to find a system where this
condition will not be the case since in conventional $SmC^{*}$ liquid
crystals $\sigma _{\perp }<1$ and $2K_3/K_2\simeq 4 - 6$ \cite{GP}. In 
contrast, the practical identification of this small-$b$ $TGB_{Ct}$ phase
can be quite difficult because of the very large cholesteric pitch. We
formulate the realistic criterion for the existence of the $TGB_{Ct}$ phase
as a condition when it is stable in the vicinity of the critical point $M_0.$
On the basis of the plot in Fig. 3c, one obtains that $t_{c2}>t_c$ near $M_0$
if:

\begin{equation}
\kappa _2>0.5\sigma _{\Vert }^{1/4}.  \label{Crit}
\end{equation}
At realistic values of $\sigma _{\Vert } \simeq 10 - 100$ \cite{Tri} 
this condition practically coincides with criterion for stability of the $%
TGB_A$ phase (\ref{CritA}). The conditions (\ref{CritA}) and (\ref{Crit})
determine the position of the critical points $M_A$ and $M_C$ on the $N^{*}$-%
$SmA,C^{*}$ transition line of Fig.1 where the $TGB_A$ and $TGB_C$ phases
first appear.

\subsection{Resume}

We have revised the calculations of \cite{Renn} for the $N^{*}$-$TGB_{A,Ct}$
phase transition. The line of the upper critical temperature on phase
diagram of Fig.1 is reconstructed on the quantitative level and can be
present in the relevant limit $\sigma _{\Vert }>1$ as follows:

On the left side of the point $M_A$ that is defined by condition (\ref{CritA}%
), the direct $N^{*}$-$SmA$ first order transition takes place. The critical
temperature of the transition is given by (\ref{TcA}).

Between the points $M_A$ and $M_0$ the transition to the $TGB_A$ phase
occurs. The point $M_0$ is placed at $0.36\sigma _{\perp }\simeq \sigma
_{\Vert }^{1/4}b$, inside the region $\sigma _{\perp }>0$. The upper
critical temperature $t_{c2}$ of the transition is given by (\ref{Tc2A})
when $-\sigma _{\perp }/b>0.5+0.38\sigma _{\Vert }$, and by (\ref{Ass2})
when $\sigma _{\perp }/b$ varies from a small negative value to $\sigma
_{\perp }/b\simeq \sigma _{\Vert }^{1/4}/0.36$. Both the asymptotics are
matched in the intermediate region of negative $\sigma _{\perp }/b$.

In between the points $M_0$ and $M_C$ the sequence of transitions $N^{*}$-$%
TGB_{2q}$-$TGB_{Ct}$ takes place, the location of the point $M_C$ being
given by the condition (\ref{Crit}). The $TGB_{2q}$ phase exists in a small
temperature interval below $t_{c2}$. It practically disappears on the right
of the point $M_1$ where $\sigma _{\perp }/b>\sigma _{\Vert }^{1/4}/0.19$.
The upper critical temperature in between $M_0$ and $M_C$ is given by 
Eq.(\ref{tc2par}). The junction of $t_{c2}$ transition lines for $TGB_A$ and 
$TGB_{2q}/TGB_{Ct}$ phases forms a kink in the tetracritical point $M_0$.

On the right side of the point $M_C$ the direct first order $N^{*}$-$SmC^{*}$
transition occurs at the critical temperature $t_c$, as is given by 
Eq.(\ref{Tc})

We have also calculated the width $\overline{l}_b$ of the $TGB$ slab
at $T=T_{c2}$ with respect to different parameters $b$,$\sigma _{\perp }$,$%
\sigma _{\Vert }$ (Eqns. (\ref{Lba}), (\ref{Lbb}), (\ref{Lbc}) and (\ref
{Lbd})). We summarize these results in  Table I by the ratio 
$l_b/l_d=\overline{l}_b/\overline{l}_d=\overline{l}_b^2b/2\pi $, 
resulting from the topological constraint (\ref{Topol}).

\section{Analogy with space-modulated superconductivity}

In this section we discuss a remarkable similarity between the $TGB$ state
and the mixed (vortex state) in superconductors of type II noted first in 
\cite{First}. We show that the $TGB_C$ phase corresponds to the mixed state
in the ''exotic'' superconductors with space modulated order parameter.

Note first that, due to de Gennes \cite{dG,GP} the $N-SmA$ transition is
analogous to the superconducting phase transition because of similarity of the
transformational properties of the order parameters: the space modulated
function $\psi (r)=\psi _0e^{q_0{\bf nr}}$ in $SmA$ and the complex wave
function $\Psi (r)$ in superconductor. The translation of $\psi (r)$ along
the modulation vector $q_0{\bf n}$ (that is equivalent to multiplication of $%
\psi $ on a phase factor) corresponds to the gauge transformation in the
superconductor.

The long range twist of ${\bf n}(r)$ in $N^{*}$ plays the role of the
magnetic field destroying superconductivity. The $N^{*}$-$SmA$ transition
via the intermediate $TGB_A$ phase occurs with continuous Meissner-like
expulsion of the twist of ${\bf n}$. The rows of the screw dislocations
resemble the Abrikosov vortex lattice.

The $SmC$ phase is characterized by the additional, transversal to ${\bf n}$
modulation of the order parameter resulted from the negative gradient terms
in the CL model. Continuing the analogy with superconductivity
one can tell that $N$-$SmC$ transition corresponds to the transition to the
superconducting state with nonuniform, space modulated order parameter. The $%
TGB_C$ phase therefore should be an analog of the mixed (Abrikosov) state of
modulated superconductor, providing that magnetic field is perpendicular to
the modulation.

On the quantitative level, the modulated superconductor in a magnetic field
is described by the GL equation:

\begin{equation}
t\psi +g\left| \Psi \right| ^2\Psi =C_{\perp }\left( i\nabla -{\bf A}\right)
^2\psi +D\,\left( i\nabla -{\bf A}\right) ^4\psi  \label{SC}
\end{equation}
where ${\bf A}$ is the vector potential of the field: ${\bf B}=curl{\bf A}$.
The modulation of the superconducting order parameter is provided by the
 gradient term with $C_{\perp }<0$. The upper
critical field of this superconductor is calculated by neglecting the
nonlinear term $g\left| \Psi \right| ^2\Psi $ and solving the corresponding
eigenproblem. It is easy to show that if ${\bf A}$ is chosen in the Landau
gauge ($0,Bx,0$), this procedure is exactly reduces to the case ''$\sigma
_{\Vert }=0$'' we considered in Sect. IIIB. According to calculations of
Sect. IIIB, this superconductor should have the oscillating upper critical
field, shown in Fig. 3a.

It is interesting to note that although the theory of space modulated superconductor
was considered by Larkin and Ovchinnikov \cite{LO} and Fulde and Ferrel \cite{FF}
more than thirty years ago, to our knowledge 
no direct experimental observation of this phase currently exists.

Note finally another interesting analogy between the $N^{*}$-$TGB_{A,Ct}$
phase transition and the $B_{c2}$ transition in ''unconventional''
multicomponent superconductor $UPt_3$ where the kink-like behavior of $%
B_{c2} $ was observed \cite{Luk}. The kink point in $B_{c2}$ in $UPt_3$ is
similar to the point $M_0$ in the $N^{*}$-$TGB_{A,Ct}$ phase diagram: both
points are provided by the intersection of eigenstates of the
corresponding linearized GL equations.

{\bf ACKNOWLEDGMENTS}

I am grateful to Monique Brunet who drew my attention to the problem and to
Laurence Navailles and Philip Barois for fruitful discussions of the
experimental aspects.  
This work was supported by the Brazilian Agency Fundacao de Amparo a Pesquisa em Minas Gerais (FAPEMIG) and by Russian Foundation of
Fundamental Investigations (RFFI), Grant No. 960218431a. Part of the work was done during my stay in  Universit\'e Montpellier II , France.

\begin{table}
\caption{Location of the   tetracritical point $M_0$ and the  ratio $l_b/l_d$ in a different limit cases. Parameter $\epsilon$ varies from $0$ when $K_{1,3}/K_2=0$ to $1.5$ when  $K_{1,3}/K_2$ is large.}
\begin{tabular}{c c c c c }
%\hline
  &\multicolumn{2}{c}{\ Location of $M_0$}& \multicolumn{2}{c}{$l_b/l_d$
\tablenotemark[1]}\\
 &$b_0$&  $t_0$ & $TGB_A$, & $TGB_{2q,Ct}$, \\ 
 & & & $b>b_0$ & $b<b_0$ \\
\hline 
$\sigma _{\Vert }\ll 1$ & $0.5\sigma _{\perp }$ & $0.75\sigma _{\perp }^2$& $0.8\varepsilon $ & $%
\sigma _{\perp }/2\pi b$ \\  
$\sigma _{\Vert }\gg 1$ & $0.36\sigma _{\perp }/\sigma _{\Vert }^{1/4}$ & $0.43\sigma _{\perp }^2$& $%
0.7\varepsilon \sigma _{\Vert }^{-1/4}$ & $\varepsilon (4\sigma _{\perp
}/b\sigma _{\Vert })^{1/3}$ \\ 
%\hline
\end{tabular}
\tablenotetext[1]{The ratio $l_b/l_d$ is given at $N^*$-$TGB$ transition line. At lower temperatures $l_b/l_d$ increases rapidly.}
%\label {table}
\end{table}

\begin{figure}[t]
\centerline{\epsffile{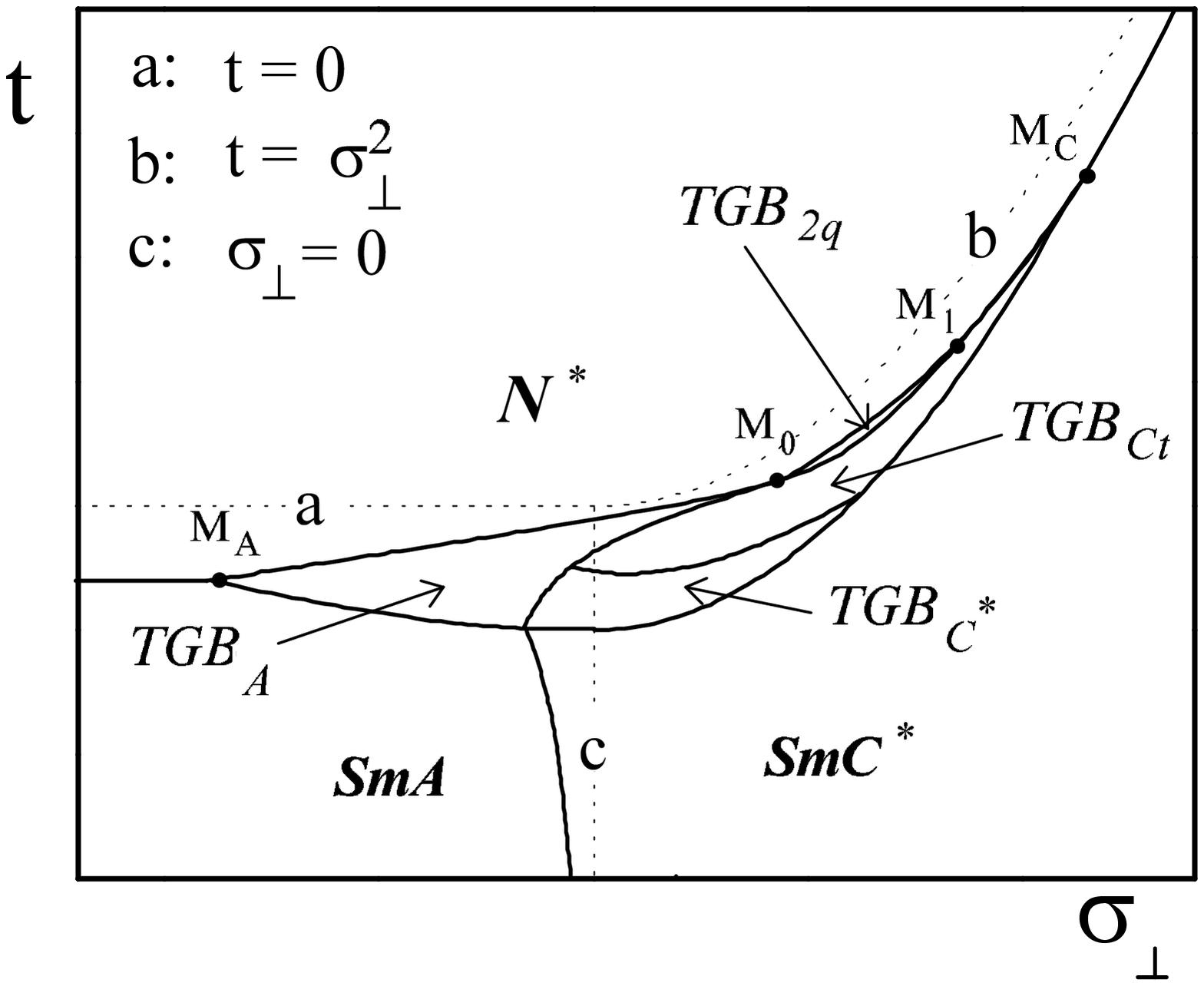}}
\caption{The phase diagram of $TGB$ phases. The model parameters 
$t,\sigma_{\bot }$ are controlled by the  experimental conditions. We predict a new $TGB_{2q}$ phase and penetration of the $TGB_A$ phase in the $SmC^*$ region where $\sigma_{\bot}>0$. Dotted lines a, b and 
c present the $N-SmA-SmC$ diagram in the nonchiral case.}
\label{fig1}
\end{figure}

%%%%%%%%%%%%%%%%%%%%%%%%

\begin{figure}[t]
\centerline{\epsffile{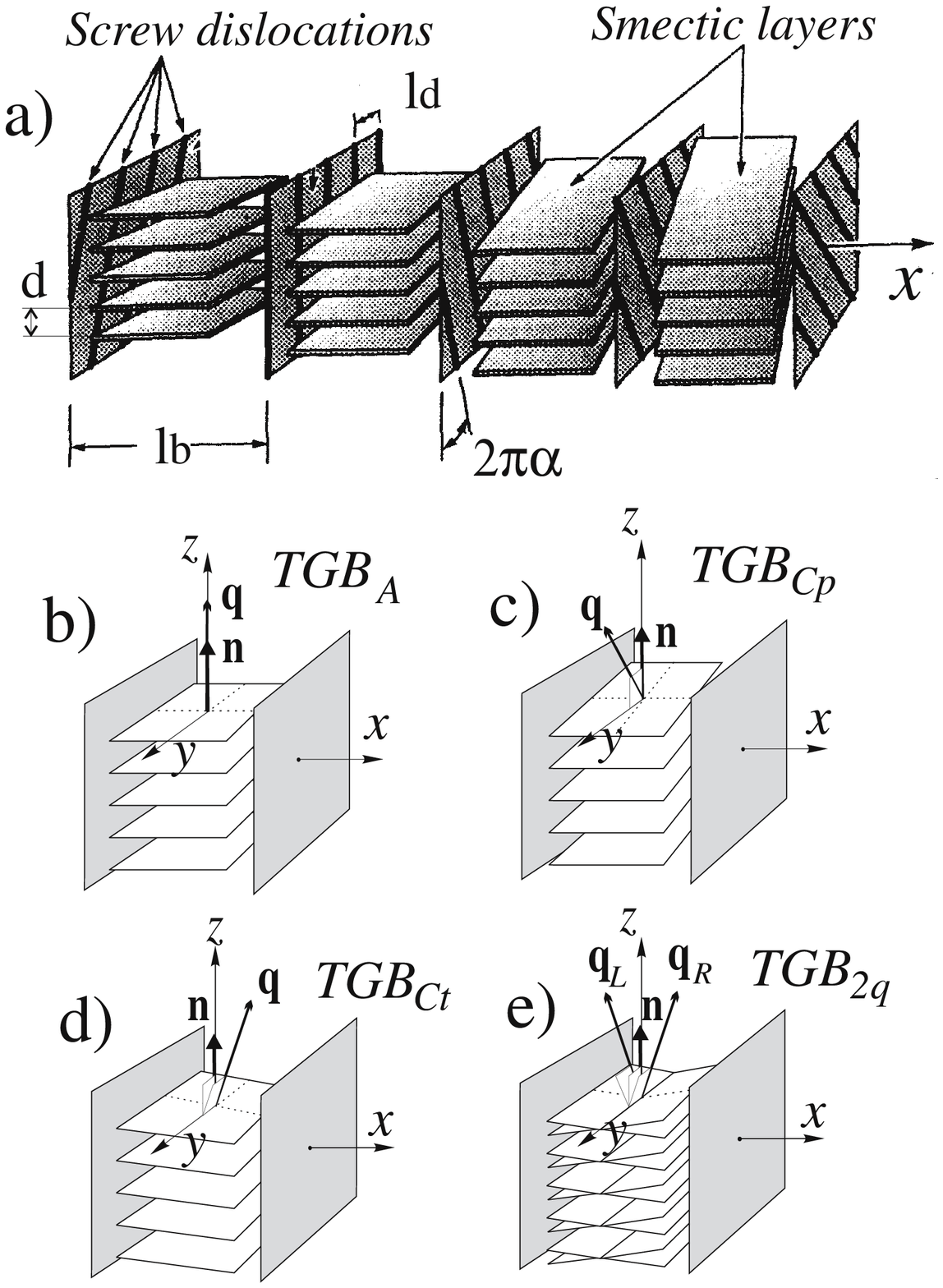}}
\caption{ Structure of different TGB phases.}
\label{fig2}
\end{figure}

%%%%%%%%%%%%%%%%%%%%%%%%

\begin{figure}[t]
  
\centerline{\epsffile{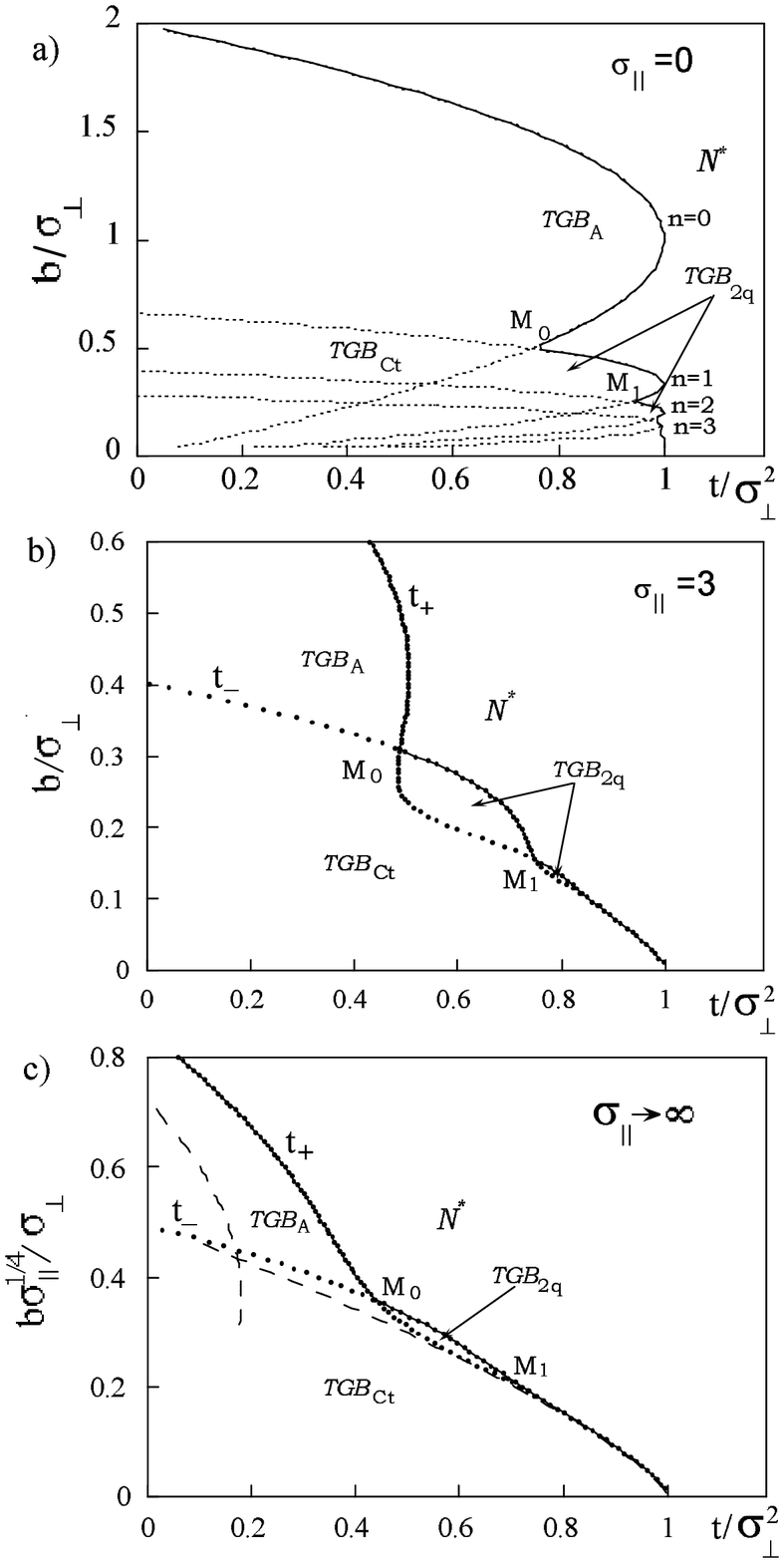}}
\caption{Oscillating behavior of the lowest eigenstates of the lineariezed CL model that correspond the Cholesteric ($N^*$) - $TGB_{Ct}$  
transition as a 
function of the model parameters: 
$t/\sigma_{\perp }^2$,  $b/\sigma _{\perp }$ and $\sigma _{\Vert }$.  
The concurrence between two nearly degenerate lowest eigenlevels (the minimal one corresponds to $t_{c2}$) leads to spliting of the transition by appearance of either the $TGB_A$ phase or the intermediate $TGB_{2q}$ phase. Dashed lines in (c) present the asymptotics calculated in the text.}
\label{fig3}
\end{figure}

%%%%%%%%%%%%%%%%%%%%%%%%

\end{document}